\documentclass{article}
\usepackage{spconf}

\usepackage{amsmath,amsfonts,bm}
\providecommand{\abs}[1]{\lvert#1\rvert}

\usepackage{graphicx}
\graphicspath{{Figures/}}
\newcommand\sfigure[4]{\begin{minipage}{#1}\centering%
\includegraphics[width=#2]{#3}\\#4\end{minipage}}
\newlength\sfigsep
\newlength\imgwidth
\usepackage{url}

\usepackage{tightbib}

\usepackage{color}

\begin{document}

\ninept

\title{Recovering Missing Coefficients in DCT-Transformed Images}

\twoauthors%
{Shujun Li\thanks{More information about this paper can be found at \texttt{http://www.hooklee.com/default.asp?t=ICIP2011}.}\thanks{Shujun Li and Andreas Karrenbauer were supported by the Zukunftskolleg, University of Konstanz as part of the ``Excellence Initiative'' Program of the DFG (German Research Foundation). The authors would like to thank Dr. Vladimir Bondarenko and Mr. Muhammad Imran Khan for their help in testing an early method based on the MATLAB Optimization Toolbox.}, Andreas Karrenbauer, Dietmar Saupe}
{Dept. of Computer and Information Science\\
University of Konstanz, Germany}%
{C.-C. Jay Kuo}
{Ming Hsieh Dept. of EE\\
University of Southern California, USA}

\maketitle

\begin{abstract}

A general method for recovering missing DCT coefficients in DCT-transformed images is presented in this work. We model the DCT coefficients recovery problem as an optimization problem and recover all missing DCT coefficients via linear programming. The visual quality of the recovered image gradually decreases as the number of missing DCT coefficients increases. For some images, the quality is surprisingly good even when more than 10 most significant DCT coefficients are missing. When only the DC coefficient is missing, the proposed algorithm outperforms existing methods according to experimental results conducted on 200 test images. The proposed recovery method can be used for cryptanalysis of DCT based selective encryption schemes and other applications.

\end{abstract}

\section{Introduction}

As a sub-optimal de-correlation transform, the discrete cosine transform (DCT) \cite{ANR:DCT:IEEETComp1974} is one of the most widely used transforms in signal and image processing applications, especially lossy audio, image and video compression. Many mainstream multimedia coding standards, such as JPEG, MPEG-1/2/4 and H.264/AVC, are built on top of DCT. To reduce computational complexity, DCT is normally performed on $N\times N$ blocks for digital images and videos. Since DCT is a good de-correlation transform, all non-zero quantized DCT coefficients in each image block are coded to allow recovering the original image/video with a desired level of visual quality. The remaining correlation between adjacent image blocks is normally handled by differential coding of DC coefficients or more complicated intra predictive coding mechanisms \cite{Shi-Sun:IVC2nd:Book2008}.

Among all DCT coefficients, the first one (called the DC coefficient) contains the most energy as well as perceptual information of an image block. Roughly speaking, the further a DCT coefficient is from the DC coefficient, the less perceptual information is represented by it. Since DCT coefficients of each image block are largely uncorrelated, all DCT coefficients of a block can be considered as separate ``quality layers'' with different significance. By exploiting this fact, some researchers proposed to implement perceptual encryption of digital images and videos by selectively encrypting part of DCT coefficients \cite{Shi:MPEGEncryption:MMTA2004, Li:PVEA:IEEETCASVT2007, Weng-Preneel:DC-Encryption:SIGMAP2007, Dufaux:VideoSE-DCT:IEEETCASVT2008}. For instance, Li \textit{et al.} \cite{Li:PVEA:IEEETCASVT2007} proposed to encrypt DC coefficients and sign bits of other DCT coefficients (called AC coefficients) to conceal the rough and detailed views of MPEG-encoded video sequences, respectively.

Since DCT is an orthogonal transform, selective encryption of DCT coefficients was believed to be secure until Uehara \textit{et al.} \cite{USO:AC2DC:IEEETIP2006} reported that missing DC coefficients can be approximately recovered from AC ones. The method proposed in \cite{USO:AC2DC:IEEETIP2006} was improved by Li \textit{et al.} \cite{LASK:FRM:ICIP2010}, which reduces propagation errors and adopts a better estimate of the first DC coefficient via under-/over-flow rate minimization (FRM). Both DC recovery methods are based on two common properties of digital images. First, there is strong correlation between neighboring pixels. Second, a smaller range of the DC coefficient of a block can be calculated from the AC coefficients of the same block. Besides the above two DC recovery methods, Said \cite{Said:PES-Measure:ICIP2005} developed a method to recover selectively encrypted AC sign bits, where a low-quality version of the original image is required to support the recovery process. None of the previous work can be easily generalized to handle the case when both the DC coefficient and some AC coefficients are encrypted. It remains an open question whether a combination of DC/AC encryption is still secure.

In this work, we propose an algorithm to recover an arbitrary set of missing DCT coefficients (except for the case when all DCT coefficients are missing). It offers a generic framework in which the DC recovery problem becomes a special case. To the best of our knowledge, this is the first time that a solution is developed to tackle such a generic and challenging problem. Simply speaking, we model the DCT coefficients recovery problem as an optimization problem and use linear programming to solve it. We have conducted experiments on 200 test images and shown that the proposed algorithm outperforms two existing DC recovery methods given in \cite{USO:AC2DC:IEEETIP2006, LASK:FRM:ICIP2010} significantly and consistently. The newly proposed algorithm does not depend on a low-quality version of the original image. Instead, it attempts to recover missing DCT coefficients solely from the information contained in known DCT coefficients. Thus, it does not suffer the limitation reported in \cite{Said:PES-Measure:ICIP2005}. Although the DCT coefficients recovery problem is studied in the context of selective encryption, the proposed method for solving the problem is actually application-independent and may find potential applications in related areas such as image/video compression and error concealment.

The rest of this paper is organized as follows. The proposed method is described in Sec.~\ref{sec:method}. Experimental results of an implementation of the proposed method are reported in Sec.~\ref{sec:experiments}, where we compare its performance with those obtained by methods in \cite{USO:AC2DC:IEEETIP2006, LASK:FRM:ICIP2010} for performance benchmarking. Concluding remarks and future research directions are given in Sec.~\ref{sec:conclusion}.

\section{Proposed Method}\label{sec:method}

In this section, we first discuss the modeling of the DC recovery problem after a brief introduction to two existing DC recovery methods. Then, its generalization to the recovery of an arbitrary set of missing DCT coefficients is addressed. After setting up the mathematical model, we will study how to solve the optimization problem effectively in terms of low space and time complexity.

\subsection{Basic Problem: DC Recovery}

The two DC recovery methods proposed in \cite{USO:AC2DC:IEEETIP2006, LASK:FRM:ICIP2010} are based on the following two properties.

\noindent
\textbf{Property 1}: \textit{The difference between any two neighboring pixels is a Laplacian variate with zero mean and a small variance.}

\noindent
\textbf{Property 2}: \textit{For each block, the range of pixel values calculated from AC coefficients constrains the value of the DC coefficient.}

Property~1 is a well-known feature of most natural images while Property~2 is a result of the relationship between the pixel values $\{x(i,j)\in[x_{\min},x_{\max}]\}_{0\leq i,j\leq N-1}$ and the DCT coefficients $\{y(k,l)\}_{0\leq k,l\leq N-1}$ as defined by $N\times N$ 2-D DCT itself:
\begin{eqnarray}
x(i,j) & = & \sum\nolimits_{0\leq k,l\leq N-1}A(i,j,k,l) \cdot y(k,l),
\label{eq:DCT2} \\
& = & \frac{1}{N} y(0,0) +
\sum\nolimits_{0\leq k,l\leq N-1 \atop (k,l)\neq (0,0)}A(i,j,k,l)y(k,l),\nonumber
\end{eqnarray}
where
\[
A(i,j,k,l)=C(k)C(l) \cos\left(\frac{(i+0.5)k\pi}{N}\right)
\cos\left( \frac{(j+0.5)l\pi}{N}\right),
\]
$C(k)=\sqrt{1/N}$ when $k=0$ and $\sqrt{2/N}$ when $k>0$. In above, the DC coefficient, denoted by $y(0,0)$, is obviously constrained by the sum term, which is DC-free, and $[x_{\min},x_{\max}]$. Note also that Eq.~\eqref{eq:DCT2} is a linear map and the indices are relative to each block. We will use $x = A \cdot y$ to denote the block-wise DCT below.

The DC recovery method in \cite{USO:AC2DC:IEEETIP2006} scans the whole image block by block and tries to align all DC coefficients to minimize the sum of differences of pixel pairs along each block boundary. Property~2 is used to estimate the global intensity of the whole image. Main drawbacks of this method include large propagation errors and a less accurate estimate of the global intensity. The improved DC recovery method in \cite{LASK:FRM:ICIP2010} introduces online pixel under-/over-flow removal in the scanning process and an under-/over-flow rate minimization process to estimate the global intensity more accurately. While the improved method can produce fairly good results for many images, the quality of recovered images is not always high. This can be explained by the fact that the scanning process is essentially local in the sense that only two neighboring blocks are considered at any time and cannot always lead to the global optimum solution.

Actually, the DC recovery problem can be formally modeled as an optimization problem w.r.t. some measure on the reconstructed pixel values and be solved to get the global optimum solution. We use variables $x(i,j)$ and $y(k,l)$ to denote the value of pixel $(i,j)$ and the DCT coefficient $(k,l)$, respectively. Then, the DC recovery problem can be written as
\begin{eqnarray}
\text{minimize} &
f \big( \{x(i,j)\}_{0 \le i,j \le N-1} \big) \nonumber \\
\text{subject to}& x = A \cdot y, \nonumber \\
& x_{\min} \le x(i,j) \le x_{\max}, \label{eq:opt}\\
& y(k,l)=y^*(k,l) \text{ for all AC coefficients.} \nonumber
\end{eqnarray}
The last equality fixes AC coefficient $y(k,l)$ to its known value $y^*(k,l)$ while DC coefficient $y(0,0)$ of each block remains to be a variable. Note that the bounds on $x(i,j)$ and the transformation constraints imply lower and upper bounds on variable $y(0,0)$.

The objective $f(\cdot)$ may be any convex function to allow a polynomial-time optimization algorithm provided that $f$ can be evaluated in polynomial time. Because of Property~1, we choose $\sum_{\{(i,j),(i',j')\}} \abs{x(i,j) - x(i',j')}$, where the sum ranges over all pairs of neighboring pixels $(i,j)$ and $(i',j')$. Moreover, this choice makes the optimization problem a linear one, which can be solved efficiently. The linearization is done by introducing variables $h_{i,j,i',j'}$ for each term of the sum and the optimization problem becomes
\begin{eqnarray}
\text{minimize} &
\sum  h_{i,j,i',j'} \nonumber \\
\text{subject to} &
  x(i,j) - x(i',j')\le h_{i,j,i',j'}, \\
& x(i',j') - x(i,j)\le h_{i,j,i',j'}, \nonumber \\
&\text{and the constraints of Eq.~\eqref{eq:opt}.} \nonumber
\end{eqnarray}

We can derive the implied bound $0\le h_{i,j,i',j'}\le x_{\max}-x_{\min}$ accordingly. Furthermore, variable $h_{i,j,i',j'}$ will be tight at $\abs{x(i,j) - x(i',j')}$ for the optimum solution.  Since the DC value of a block contributes to all pixels of the same block equally, only the pairs of neighboring pixels that belong to different blocks are relevant. Hence, we may restrict ourselves to these pairs.

The above optimization problem contains a free variable corresponding to the global intensity when the dynamic range of the optimum solution is smaller than $[x_{\min},x_{\max}]$. In this case, the global intensity can be shifted without changing the value of the objective. In other words, there are multiple optimum solutions. We handle this problem by shifting the histogram of the recovered image (i.e., one solution) towards the midpoint of $[x_{\min},x_{\max}]$ until the left and right margins of the histogram are equal.

\subsection{General Problem: Arbitrary DCT Coefficients Recovery}

The generalization from DC recovery to arbitrary DCT coefficients recovery is straightforward. That is, we can fix $y(k,l)=y^*(k,l)$ for known DCT coefficients while constrain all remaining unknown variables in a range $[y_{\min}(k,l), y_{\max}(k,l)]$ depending on their locations ({\em i.e.} spatial frequencies). Note that when at least one AC coefficient is unknown, pixel pairs inside each block shall \emph{not} be discarded as done in the previous case. The above generalization clearly maintains the linearity of the optimization problem. Hence, we can use the \emph{linear programming} (LP) technique to solve both the basic DC recovery problem and the general DCT coefficients recovery problem. In our implementation, we adhere to the general problem and treat the DC recovery problem as a special case.

\subsection{Solving Optimization Problem}

From a theoretical point of view, linear programs can be solved in time bounded polynomially in the binary encoding length of the input data using the \emph{ellipsoid method}~\cite{Khachiyan79} or the \emph{interior point methods}~\cite{Karmarkar84}. The latter yields an efficient implementation in practice. Other methods with good practical performance include the primal and dual \emph{simplex} algorithm. We refer to \cite{BT:LinearOptimization:Book97} for interested readers.

The LP solution consists of floating point values for the pixel values and DCT coefficients. After the optimization, we round the pixel values to fixed precision numbers, i.e. to integers. Our experimental results show that the rounding error is negligible in terms of visual quality, so we do not re-optimize after the rounding step. Additional integrality constraints for pixel values lead to \emph{Integer Linear Programming} (ILP), which is NP-hard in general \cite{Karp:NPC-List:1972}. Whether this also holds for our application is a topic for further research. As of yet, we know that the problem remains polynomial time solvable for the simpler case of DC-only recovery. The reason for this is that there is an equivalent formulation as an LP over an integer polyhedron for which ILP is not more difficult than LP \cite[Chapter~5.15]{Schrijver03}.

\subsection{Complexity Analysis}\label{section:Complexity}

Given an $n\times m$ image and $U$ unknowns in each of $B$ $N\times N$ blocks, the mathematical model presents a large-scale optimization problem with $2nm-(n+m)$ $h$-variables and $U\cdot B$ $y$-variables. Since all other unknowns can be uniquely determined by these variables, they can be eliminated by a presolve step of the optimization process. For the basic DC recovery problem, the number of variables drastically reduces as only pixel pairs along block boundaries contribute to the objective value. Assuming that $n$ and $m$ can be divided by $N$, the number of variables involved is reduced to $nm/N-(n+m)$ $h$-variables and $nm/N^2$ $y$-variables.

The worst-case time complexity of the interior point method for solving LP problems with $V$ variables and with a size of $L$ bits is $O(V^3L)$ arithmetic operations~\cite{Ye91}. Here, the problem size $L$ is the total number of bits needed to store the whole LP including the objective, the constraint matrix and the right-hand side of the constraints. For our problem, the value of $L$ is $O(nmU)$ thanks to the sparsity of the constraint matrix. As a result, the worst-case time and space complexities are $O( n^4m^4U)$ and $O(nmU)$, respectively. Since our problem is rather sparse, the running time scales much better in practice as reported in the next Section.

When the image size is large, the required memory and computational time is untractable for low-end computers such as PCs. This problem can be partly solved by dividing the whole image into sufficiently small patches and the global intensity of different patches can be adjusted to minimize the discontinuity along patch boundaries. It ia also possible to limit the scale of the problem by reducing the number of pixel pairs used in the model. Both measures will inevitably compromise the visual quality of the recovered image. In this paper, we only consider images that can be directly handled by the optimization algorithm. The above ``divide and conquer'' idea will be a future research topic.

\section{Experimental Results}\label{sec:experiments}

To validate the real performance of the proposed method for recovering multiple missing DCT coefficients, we built an implementation based on the commercial optimization software package IBM ILOG CPLEX 12.2~\cite{IBM_CPLEX}. More precisely, we used the constrained \emph{barrier optimizer} (an interior point method) in CPLEX because it turned out to be the most suitable method for our purpose.

We ran the proposed method on 200 test images and evaluated the subjective and objective visual quality of the recovered images against the original ones to see how well the proposed method works. The objective visual quality metrics include PSNR and nine other ones included in the MeTriX MuX Visual Quality Assessment (VQA) Package \cite{MeTriXMuX}. All the 200 test images are 8-bit gray-scale images, so $x_{\min}=0$ and $x_{\max}=255$.

\subsection{Performance of DC Recovery}

\begin{figure}[!htb]
\centering
\sfigure{\imgwidth}{\imgwidth}{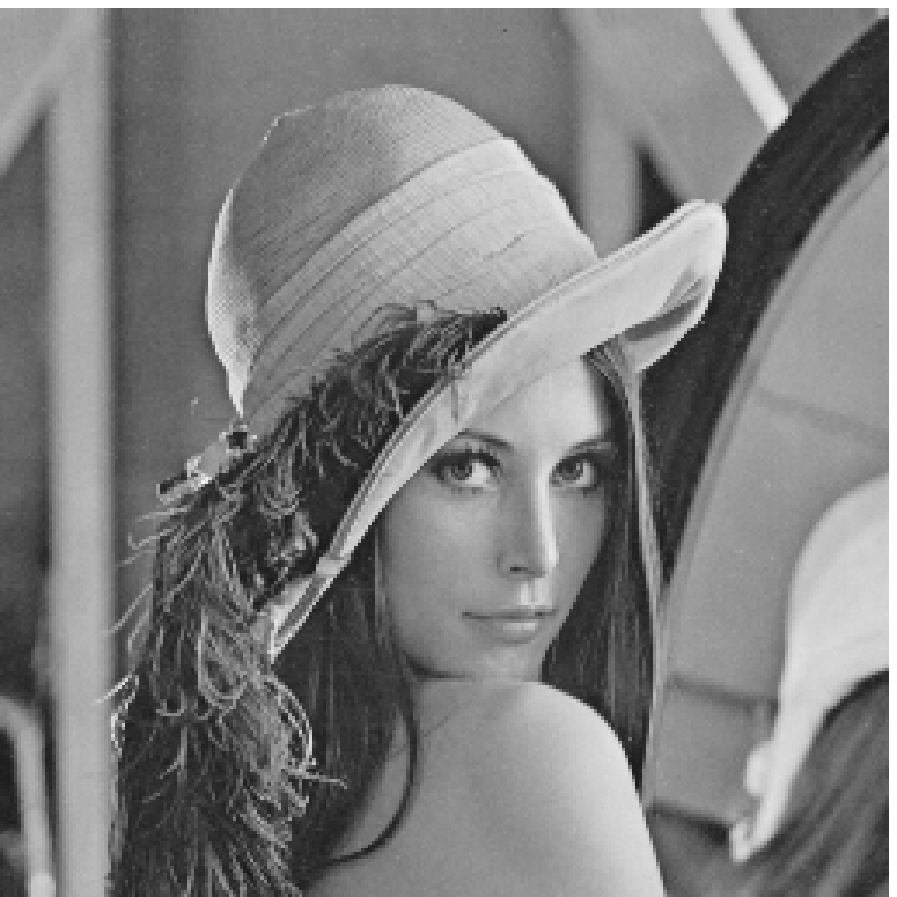}{(a)}\hfil
\sfigure{\imgwidth}{\imgwidth}{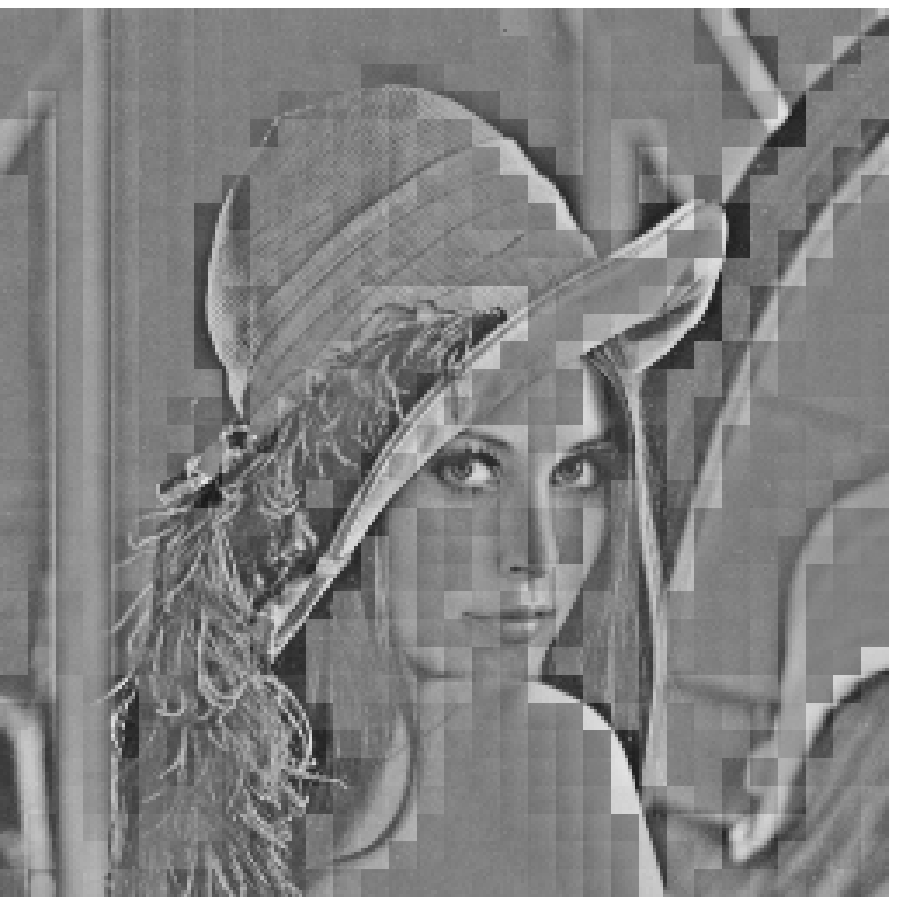}{(b)}\hfil
\sfigure{\imgwidth}{\imgwidth}{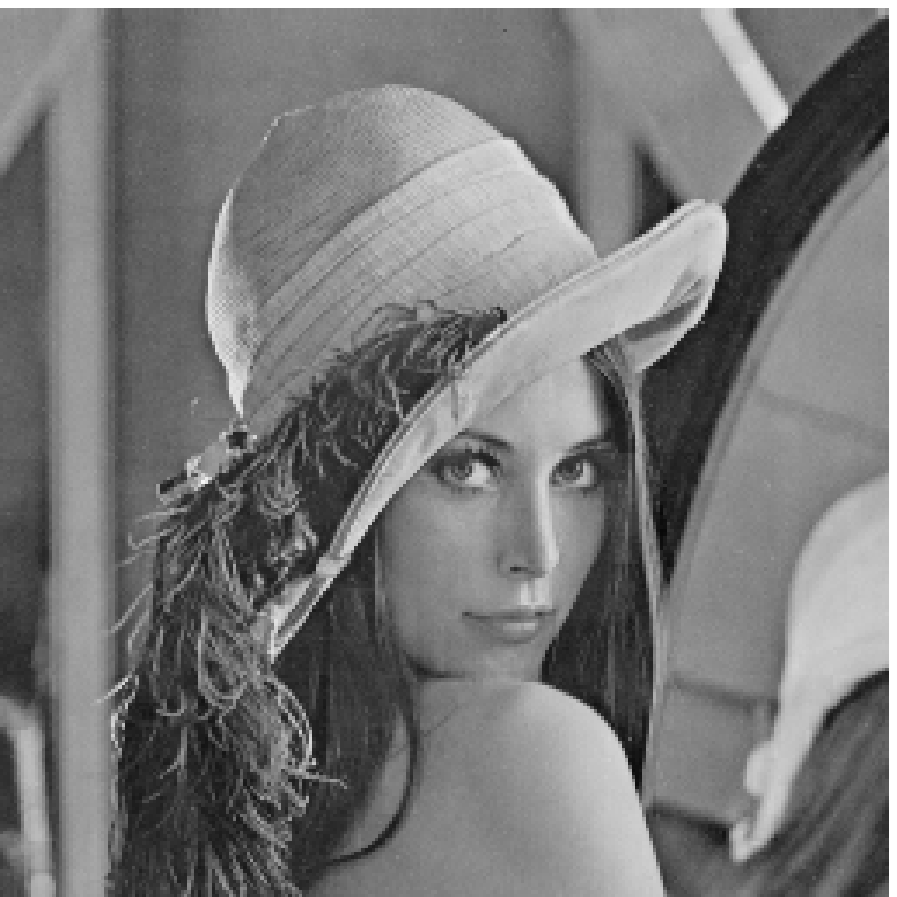}{(c)}\\
\sfigure{\imgwidth}{\imgwidth}{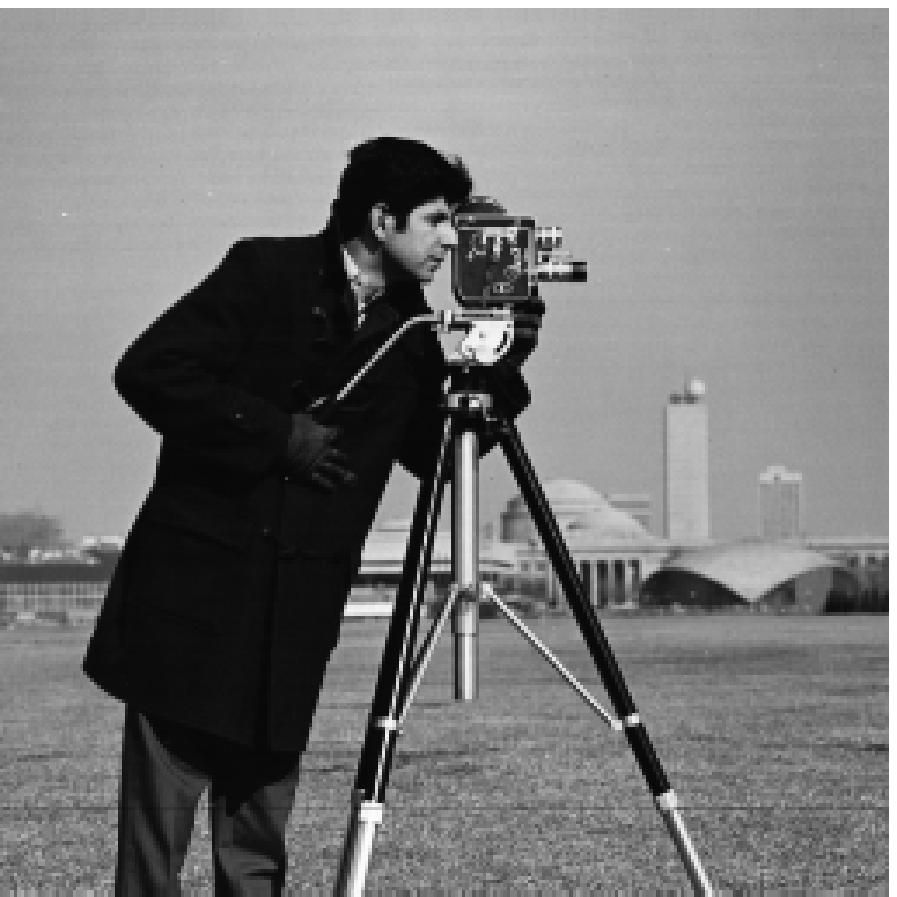}{(d)}\hfil
\sfigure{\imgwidth}{\imgwidth}{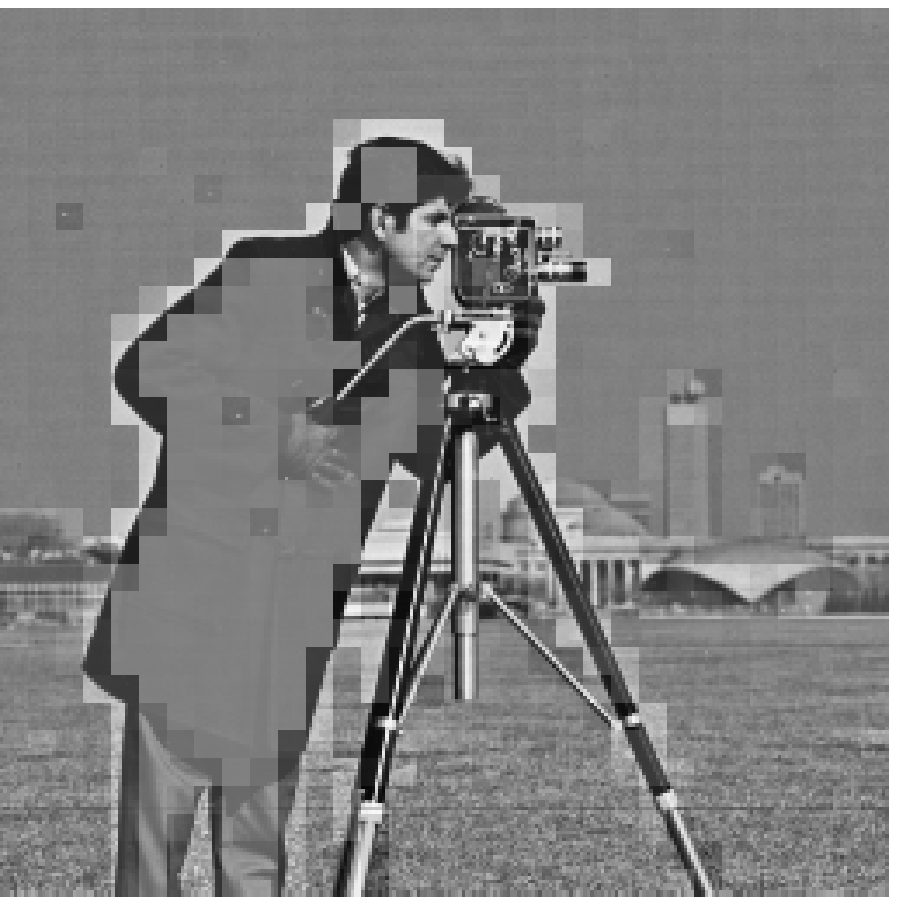}{(e)}\hfil
\sfigure{\imgwidth}{\imgwidth}{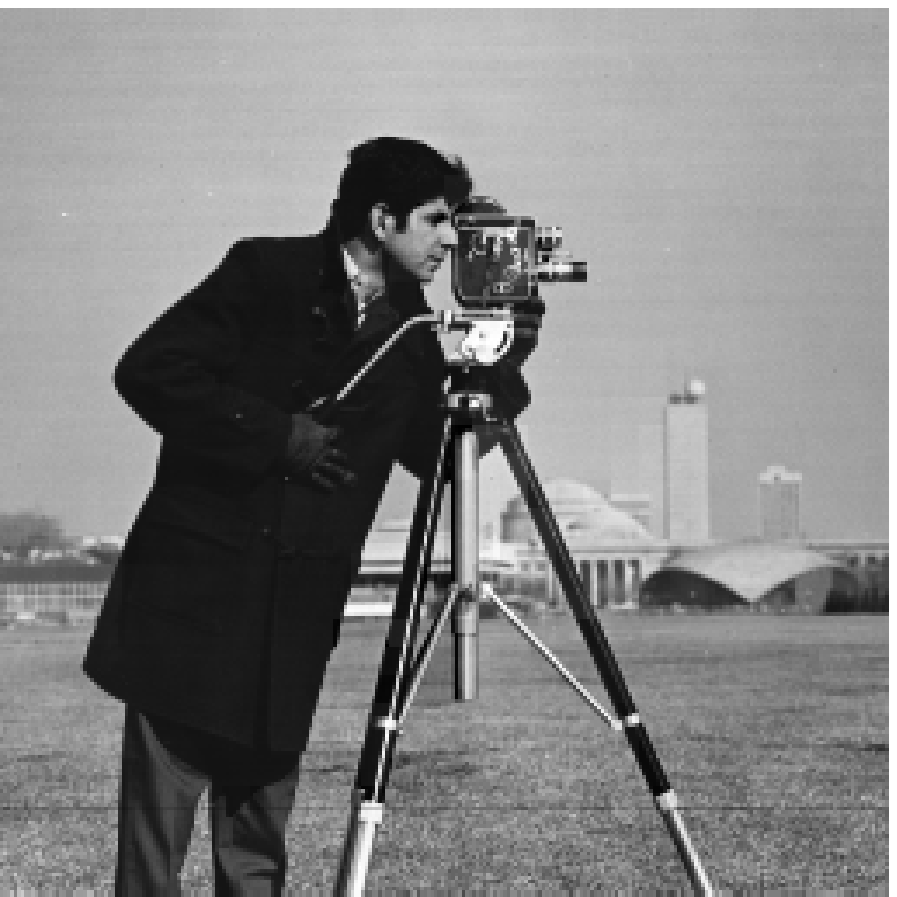}{(f)}
\caption{The recovery results of two test images (a) ``Lenna'' and (d)
``cameraman'': (b) and (e) DC-free images by setting the missing DC
coefficients to the midpoints of the valid ranges; (c) and (f)
recovered images by the proposed method.}
\label{figure:Lenna_cameraman_images}
\end{figure}

The performance of DC recovery is nearly perfect as judged the authors as human observers. For all 200 test images, we cannot see any obvious quality degradation other than some global intensity shift, which is rarely viewed as visual distortion since no perceptual information is lost. The recovery results on two typical test images ``Lenna'' and ``cameraman'' are shown in Fig.~\ref{figure:Lenna_cameraman_images}. The DC recovery process is also very efficient. The whole process could finish in less than 10 seconds on all laptop/desktop PCs we tested in our experiments and the average running time over the 200 test images is around 5 seconds on our main test computer (a quad-core desktop).

\begin{figure}[!htb]
\centering
\includegraphics[width=0.495\columnwidth,clip]{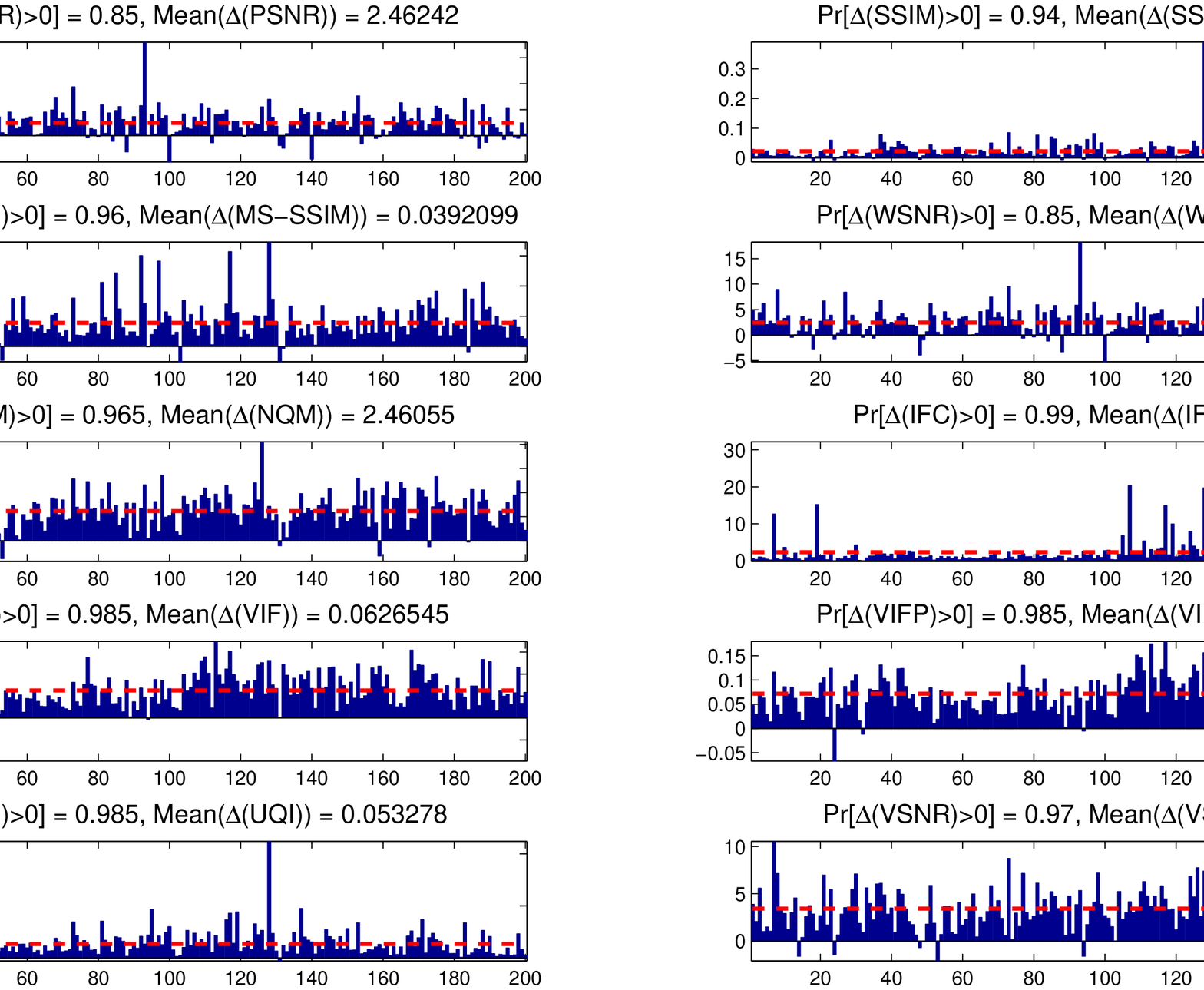}
\includegraphics[width=0.495\columnwidth,clip]{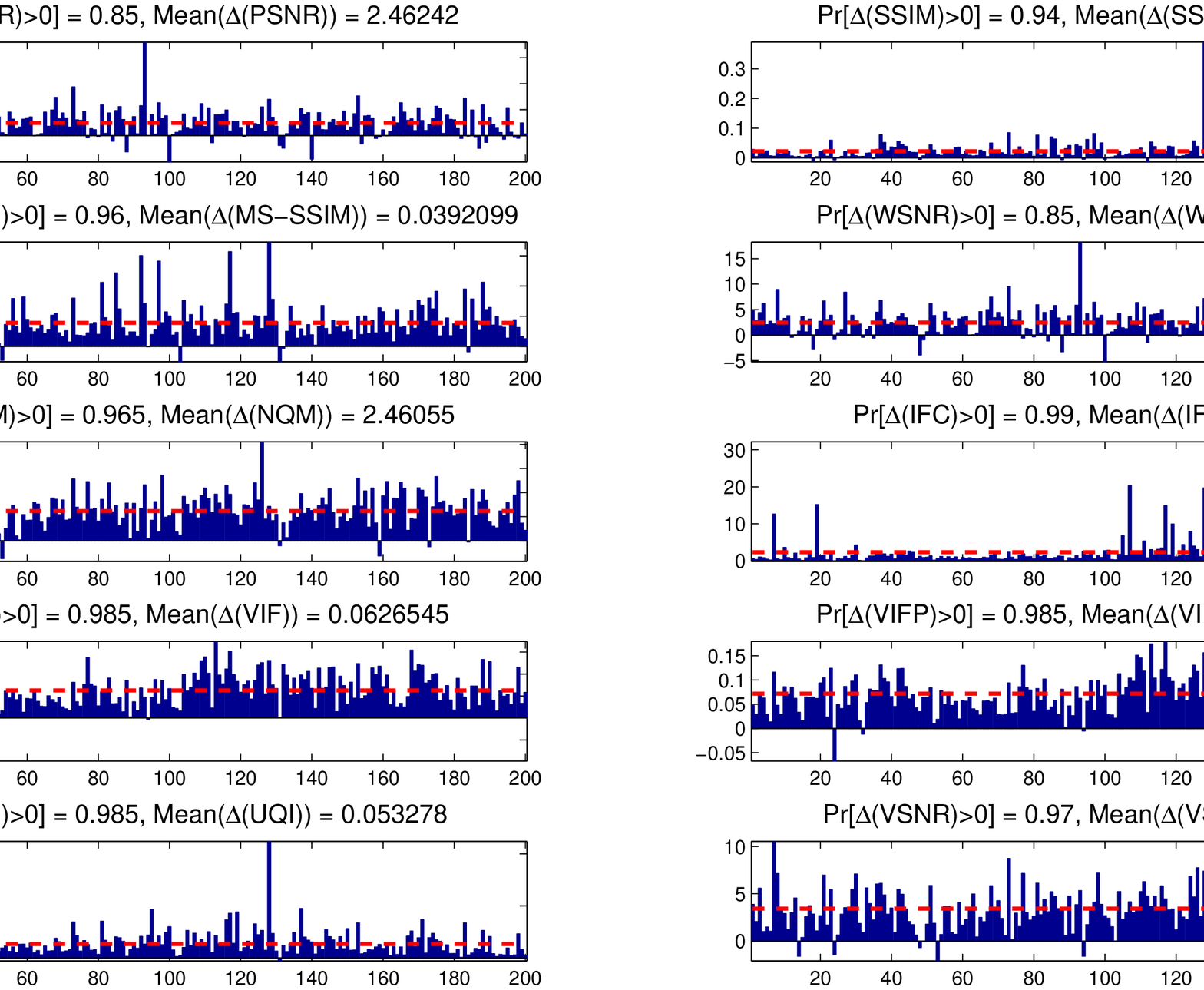}
\caption{The performance comparison between the proposed method and the DC recovery method proposed in \cite{LASK:FRM:ICIP2010}, where the red dashed line shows the mean value of the difference of quality metrics.}\label{figure:DC_Opt_FRM}
\end{figure}

We compared the performance of the proposed method and the FRM method proposed in \cite{LASK:FRM:ICIP2010}. The results are shown in Fig.~\ref{figure:DC_Opt_FRM}, where the x-axis denotes the image index and the y-axis shows the difference of quality metrics of the two methods. A positive value means that the proposed method performs better. One can see that the proposed method outperforms the FRM method significantly and consistently over the whole image database. For some VQA metrics, a few images seems to have worse quality, but a manual check by the authors revealed no noticeable difference in subjective quality. Since the FRM method outperforms the method proposed by Uehara \emph{et al.} in \cite{USO:AC2DC:IEEETIP2006}, the proposed method is the best among the three.

\subsection{Performance of DCT Coefficients Recovery}

\begin{figure}[!htb]
\centering
\includegraphics[width=\columnwidth,clip]{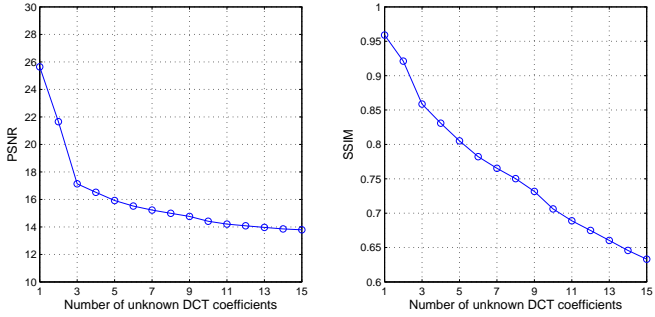}
\caption{The average visual quality measured in terms of PSNR and SSIM of the 200 test images for $U=1$, $\cdots$, 15.}\label{figure:quality_vs_U}
\end{figure}

\begin{figure}[!htb]
\centering
\sfigure{\imgwidth}{\imgwidth}{cameraman}{(a)}\hfil
\sfigure{\imgwidth}{\imgwidth}{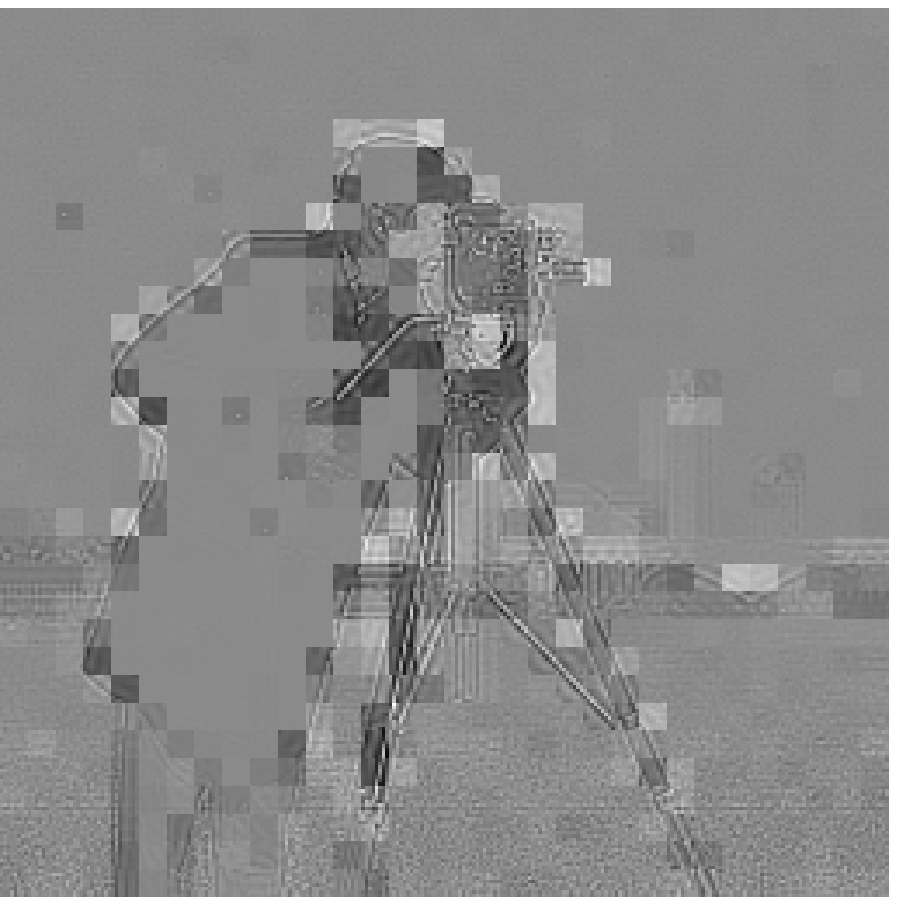}{(b)}\hfil
\sfigure{\imgwidth}{\imgwidth}{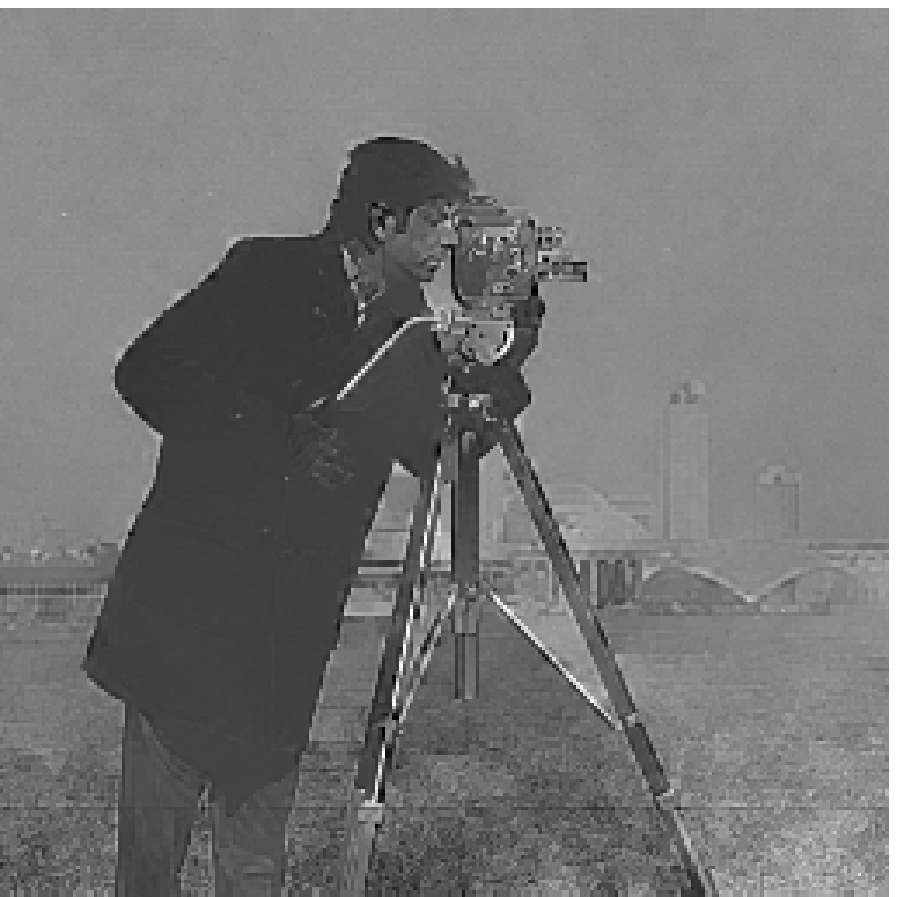}{(c)}
\caption{The recovery results when 15 most significant DCT coefficients are missing: (a) the original image ``cameraman''; (b) the image by setting all missing DCT coefficient to the midpoints of their valid ranges; (c) the recovered image by the proposed method.}\label{figure:cameraman_U15}
\end{figure}

We also studied the performance of the proposed method when $U\geq 1$ most significant DCT coefficients in each block are unknown. As shown in Fig.~\ref{figure:quality_vs_U}, the visual quality of the recovery image gradually decreases while the value of $U$ increases. This is expected since we have less and less known DCT coefficients available for recovering the missing one. The recovery performance is still good for some images even when more than 10 most significant DCT coefficients are missing as shown in Fig.~\ref{figure:cameraman_U15}). The average running time ranges from around 5 seconds ($U=1$) to around 10 minutes ($U=15$).

\subsection{Actual Complexity}

We also made some experiments to check the actual time and space complexities of the optimization problem. To make the estimated time and memory usage more accurate, we limited the number of threads to 1 although parallelism is supported by CPLEX. Note that this does not affect the reconstructed images. When the number of unknowns $U$ is fixed, the running time grows almost quadratically and the memory usage scales linearly in the number of pixels $nm$. For example, about 4-10 KB memory per pixel is used on average for $U=1$, $\cdots$, 15. When the number of pixels $nm$ is fixed, both the running time and memory usage grows linearly in the number of unknowns $U$. Therefore, we conclude that the actual time complexity is $O(n^2m^2U)$ and the actual space complexity is $O(nmU)$.

\section{Conclusion and Future Work}\label{sec:conclusion}

A method to recover multiple missing DCT coefficients in DCT-transformed images was proposed in this work. It was formulated as an optimization problem that can be effectively solved by linear programming. Experimental results validated the excellent performance of the proposed method. Besides its direct application to cryptanalysis of selective encryption schemes, the fact that $U\geq 1$ DCT coefficients can be approximately estimated from other known coefficients drives us to re-think how much information of DCT coefficients is needed in encoding and transmission.

In Sec.~\ref{section:Complexity}, we mentioned one future research topic on how to overcome the problem with large images. There are many other topics that deserve further investigation. They are given below.
\begin{itemize}
\item
More theoretical analysis about solvability, complexity of the problem and real performance of the proposed method.

\item
Further improvement of the method to balance time/space complexity and the visual quality of the recovered image.

\item
Generalization of the method to handle unknown sign bits of some DCT coefficients.

\item
Generalization of the method to permuted DCT coefficients (i.e., when locations of some DCT coefficients are unknown).

\item
Generalization of the method to color image and video by exploiting more correlation among different color channels and different video frames.

\item
Generalization of the method to other transforms, especially DWT used in JPEG2000.

\item
Applications of the method to image and video compression by coding less information about transform coefficients.
\end{itemize}

\bibliographystyle{IEEEbib}
\bibliography{ref}

\end{document}